\documentclass[12pt,preprint]{aastex}

\shorttitle{Sigma Ori IRS1 A and B}
\shortauthors{Hodapp,Iserlohe,Stecklum,Krabbe}

\begin{document}

\title{$\sigma$ Orionis IRS1 A and B: A Binary Containing a Proplyd}
\author{Klaus W. Hodapp\altaffilmark{1}, Christof Iserlohe\altaffilmark{2}, 
Bringfried Stecklum\altaffilmark{3}, Alfred Krabbe\altaffilmark{2}} 

\altaffiltext{1}{
Institute for Astronomy, University of Hawaii,\\
640 N. Aohoku Place, Hilo, HI 96720, USA
\\email: hodapp@ifa.hawaii.edu }

\altaffiltext{2}{
Deutsches SOFIA Institut, Universit{\"a}t Stuttgart,
Pfaffenwaldring 31, D-70569 Stuttgart, Germany\\
}

\altaffiltext{3}{
Th{\"u}ringer Landessternwarte Tautenburg, Sternwarte 5,
D-07778 Tautenburg, Germany\\
}

\begin{abstract} 

We report optical and infrared imaging spectroscopy observations of the young binary
object $\sigma$~Orionis~IRS1~A/B. The brighter component ($\sigma$~Ori~IRS1~A) of this binary
system has M1 spectral type and a mass in the range of $\approx$~0.3~--~0.8~$M_\odot$.  
The fainter component ($\sigma$~Ori~IRS1~B) has a unique
morphology and spectrum. The unresolved stellar object is surrounded by
an extended envelope that is slightly offset from the position of this star.
The envelope's spectrum shows strong emission lines
of H and HeI but no shock-excited emission from H$_2$ or [FeII].
The embedded stellar object $\sigma$~Ori~IRS1~B
has an absorption spectrum characteristic of a late M photosphere, but
with an additional
approximately equal amount of dust continuum flux veiling the absorption lines. 
$\sigma$~Ori~IRS1~B is probably a young brown dwarf embedded in a
proplyd that is being photo-evaporated by the UV flux of the nearby
multiple O and B star system $\sigma$~Ori.

\end{abstract}


\section{INTRODUCTION}

The binarity of the late O-star $\sigma$~Ori~was discovered by \citet{Struve1837} using visual
micrometer measurements. With the further discovery by \citet{Bolton1974} 
that component $\sigma$~Ori~A is itself a spectroscopic binary, the determination of the
orbit of components A and B by \citet{Heintz1997} and the identification of the massive C, D, and
E components \citep{Caballero2007a} and \citep{Sherry2008}, the $\sigma$~Ori system is now viewed as somewhat analogous to
the Trapezium system of massive stars in the Orion Nebula. 
The multiple star system $\sigma$~Ori lies in the center of a small cluster of 
young lower mass stars, as first discussed by \citet{Garrison1967}.
This cluster has $\approx$~300 members with
a total mass of 150 -- 225~M$_\odot$, with ages of individual members ranging
from several million years \citep{Sherry2008} down to stars in the outflow phase \citep{Reipurth1998}
and
masses ranging from the massive members of the $\sigma$~Ori
group down to objects of possibly only planetary mass \citep{Zapatero2008} (and references therein).
The properties of the $\sigma$~Ori cluster have recently been reviewed
by \citet{Walter2008} and we refer to references therein for more details.
\citet{Sherry2008} arrived at a main sequence-fitting
distance of 420 pc, consistent with the distance to 
main Orion OB1b association. The extinction towards the $\sigma$ Ori cluster is
very low and can be neglected for work in the near infrared.

The discovery by \citet{vanLoon2003} of the infrared object $\sigma$~Ori IRS1 
about 3$\arcsec$ north of the high-mass binary $\sigma$~Ori A/B,
associated with extended infrared emission and coinciding with a VLA radio source
\citep{Drake1990} raised the possibility that this object
is a proplyd, similar to the many objects of this type found in the Orion nebula.
$\sigma$~Ori IRS1 has been detected in X-rays by \citet{Caballero2007b} and \citet{Skinner2008}. The variability
and the high temperature component in its X-ray spectrum are consistent with it containing a magnetically
active low mass T Tauri star.
Infrared images using adaptive optics by \citet{Bouy2009} showed that $\sigma$~Ori~IRS1
is a double source with components IRS1~A and IRS1~B.
In this paper we present near-infrared integral-field spectroscopy of the two components
of $\sigma$~Ori~IRS1.

\section{OBSERVATIONS AND DATA REDUCTION}

We had noticed the binary nature of $\sigma$~Ori~IRS1 independently from \citet{Bouy2009}
during observations on 2007, December 30 (UT) using
the Keck 2 telescope with the NIRC2 near-infrared camera and natural guide star adaptive
optics under non-photometric conditions.
The image in the Br$\gamma$ filter is shown as part of Fig.~1. 
We observed $\sigma$~Ori~IRS1 again
with the OH-Suppressing Infrared Integral Field Spectrograph (OSIRIS) \citep{Larkin2006} 
at Keck 2 on the night of 2008, Aug. 21, UT, under excellent weather conditions
using the finest scale of 20 mas pixel$^{-1}$.
The quality of these observations allowed
a detailed study of the $\sigma$~Ori~IRS1~B emission spectrum in the {\it H} and {\it K} windows (Figs.~2 and 3) and the 
spectral classification of both components of the binary star system
$\sigma$~Ori~IRS1 (Figs.~4).
For the {\it K}-band observations, we used spectra of the multiple O and B star system $\sigma$~Ori~A/B
as the atmospheric absorption standard, after interpolation across the strong Br$\gamma$ absorption line,
and multiplied the ratio spectra by the spectrum of a 
30750 K blackbody, the T$_{eff}$ of an O9.5V star \citep{Massey2005}. 
We did not obtain calibration spectra of $\sigma$~Ori~A/B in the {\it H}-band. Therefore,
the {\it H}-band spectrum of $\sigma$~Ori~IRS1~B in Fig.~2 used the M1 star $\sigma$~Ori~IRS1~A as the atmospheric
absorption standard, and the ratio spectrum was multiplied by the spectrum of the M1V star HD42581
from the spectral library of \citet{Cushing2005}, to establish the proper continuum.
The spectrum in Fig.~2 was separately flux calibrated in the {\it H} and {\it K} bands by multiplying the
spectrum with the 2MASS spectral response curves, integrating the flux, using the 2MASS
flux calibration of \citet{Cohen2003}, and then scaling the result
to match the published {\it H} and {\it Ks} magnitudes of \citet{Bouy2009}.

We also obtained optical seeing-limited slit-less spectroscopy using the Wide Field Grism Spectrograph 2 (WFGS2) 
\citep{Uehara2004} at the UH 2.2m telescope in an effort to detect the H$\alpha$
line from $\sigma$~Ori~IRS1~B. The images were obtained on the night of 2008, October 11, UT,
under near-photometric conditions. This wide-field collimator-camera
system was used without a slit, with a narrow-band H$\alpha$ filter (10 nm FWHM) as a spectral
selector, and with the high-resolution grating, 
resulting in a dispersion of 0.085 nm pixel$^{-1}$.
As a result, all stars
are dispersed into short continua, covering the transmission range of the H$\alpha$ filter.
The central area of the resulting image is shown in Fig.~1.
The emission-line dominated $\sigma$~Ori~IRS1~B is easily seen in the resulting slit-less
spectroscopy image as the only object near the $\sigma$~Ori system that
appears as an unresolved dot of H$\alpha$ emission, without detectable continuum
or velocity structure. 

\section{SPECTRAL CLASSIFICATION OF IRS1 A AND B}

The median age of the $\sigma$ Ori cluster is 2-3~Myr \citep{Sherry2008}, so that all low mass objects
are still contracting, and are therefore larger, more luminous, and have lower
surface gravity than main sequence stars.
When attempting to classify their spectra, it is usually found that their features
fall between those of luminosity class V and III, but closer to class V, as was shown in the early work
by \citet{Hodapp1993} and \citet{Greene1996}.
In mid to late M spectral types, the ratio of the strength of the CO bandheads
to that of the atomic lines, as well as the ratio of the two components of
the Na doublet (see Fig. 4), are luminosity dependent and the latter show that both
$\sigma$ Ori IRS1 components are indeed much closer to luminosity class V than to class III. 

The spectral classification of the two components of $\sigma$ Ori IRS1 relies
on the strongest molecular feature in the atmospheric {\it K} window,
the CO bandheads, and on the atomic line systems of Na , Ca, Mg, and the complex of Fe and Ti lines
at 2.23~$\mu$m.
The spectral classification was done by discussion of features in direct
comparison (Fig.~4) with the IRTF/SPEX spectral library of 
\citet{Cushing2005}.

\subsection{$\sigma$ Ori IRS1 A}
In the spectrum of IRS1 A in Fig.~4 the absorption line systems of Na and Ca are
of almost equal strength. The absorption line of Mg at 2.282 $\mu$m is clearly detected.
This is most closely matched by the comparison
spectrum of M1V type that also reproduces the shape of the Fe and Ti complex at 2.23 $\mu$m
best. For later spectral types (M3V), the Mg line is too weak and the shortest wavelength component
of the Ca feature becomes too strong. For even earlier spectral types, Mg becomes too strong,
and the line ratios of the Ca and the Fe and Ti multiplets do not match $\sigma$ Ori IRS1 A.
The CO bandheads in IRS1 A are stronger compared to Na and Ca than in M1V
spectral types, as it would be expected for an object of higher than
main sequence luminosity. 
We conclude that a spectral type of M1V is the closest match to the spectrum
of IRS1 A with an uncertainty of one decimal sub-class, 
the caveat being that IRS1 A is more luminous than a main sequence object
and that it is young. 
Following \citet{Luhman2003}, this gives its effective temperature
as 3705~$\pm$~150~K.

From the K$_s$ magnitude of 10.48 \citep{Bouy2009} and with colors and
bolometric corrections from \citet{Johnson1965}, we arrive at a rough estimate for the luminosity
of $\sigma$ Ori IRS1 A of 0.86~$L_\odot$ .
Comparison of these T$_{eff}$ and L data with the 1998 revisions of the \citet{DAntona1997} evolutionary tracks
in Fig.~5 places $\sigma$ Ori IRS1 A on the track of a 0.3 M$_\odot$ star with a nominal age of
0.3 Myrs. 
Compared to the 1 Myr isochrone of the \citet{Baraffe1998} and \citet{Chabrier2000} models
$\sigma$ Ori IRS1 A appears to be a star of 0.8 M$_\odot$ and an age 
of $\approx$1 Myr.
The errors indicated in Fig.~5 represent the uncertainty of $\pm$1 subclass in spectral
type determination, and the estimated total error from the uncertainty in bolometric
corrections values \citep{Malkov1997}, color transformations, and photometric errors.
As discussed both by \citet{DAntona1997} and 
\citet{Baraffe2002}, the evolutionary models themselves are still quite uncertain
at these young ages,
adding another caveat to the interpretation of the results.

\subsection{$\sigma$ Ori IRS1 B}

The best fit to the line ratios of $\sigma$~Ori~IRS1~B
is a M7 or M8 star on the basis of the detection of Na lines and non-detection
of Ca lines, and the ratio of Na to CO bandhead strength. However, all these late M class V stars
have much stronger lines (both the Na and the CO bandhead lines) than $\sigma$~Ori~IRS1~B, so in order to match the
spectrum, we have to postulate an additional, diluting continuum component
to the spectrum. 
As we will discuss in detail in Section 4. the strong spatially extended 
Br$\gamma$ and HeI emission suggests that $\sigma$~Ori~IRS1~B
is a proplyd in the process of being photo-evaporated by UV radiation from the OB stars in the $\sigma$~Ori~A/B system.
It is therefore reasonable to assume that some dust in the proplyd gets heated by UV absorption
to temperatures near the sublimation point, and that its emission
dilutes the photospheric emission from the embedded star. 
However
Fig.~3 clearly shows that the line emission is more extended than the
continuum, which appears unresolved.
It is also possible that part of the hot dust emission originates much closer to the star 
and is powered by an accretion disk.
Our data are not sufficient to uniquely identify a combination of dust
opacity and temperature to best match the spectrum, but solutions with dust temperatures
in the range of 1000 K and contributions of about 50 \% of dust emission to the total
flux give good matches to the line depths.
Fig.~4 therefore compares the spectrum of $\sigma$~Ori~IRS1~B, expanded by a factor of 2 to account
for the dilution of the lines, with archival spectra of mid-M stars (above) and
M8-L1 stars and an M7III star below. 
This comparison places the spectral type of $\sigma$~Ori~IRS1~B in the range from M6 to M9.
A spectral type as early as M6 is consistent with the data when we consider that for
lower gravity atmospheres, the Na is fainter and the CO bandheads are stronger than for
main sequence stars. On the other extreme, the M9 class is still compatible with the data,
considering the substantial errors. A spectral type of M5 can be excluded with confidence,
since the Ca feature would be much too strong, and on the other side, a L1 or later type
can be excluded since those types do not show Na absorption. 
We therefore nominally classify $\sigma$~Ori~IRS1~B as M7.5, i.e., the midpoint of the
range of possible spectral types.
We assume as the effective temperature of $\sigma$ Ori IRS1 B the 
average of the effective temperatures given for M7 and M8 in \citet{Luhman2003}: 2795 K
with a range of uncertainty between 2400 K (M9V) and 3000 K (M6V).
While it is clear from its luminosity and age that $\sigma$~Ori~IRS1~B is not a
main sequence star, its spectrum does not show, within the limitations of the
signal-to-noise ratio, any specific features related to lower surface gravity.
Nevertheless, it should be classified as luminosity class IV.

With a measured {\it K$_s$}=12.65 from \citet{Bouy2009}, we obtain an M$_{bol}$ of 7.10
using colors and bolometric corrections from \citet{Johnson1965} 
and a luminosity of L = 0.109~L$_\odot$. The errors indicated in Fig.~5 represent the uncertainty 
in the determination of the spectral type, and the maximum effect of a factor of 2 in
luminosity from either including or excluding the dust emission in the determination of
the luminosity. 
It places $\sigma$~Ori~IRS1~B
on the evolutionary track of a 0.10 $M_\odot$ object in the 1998 revisions of
the \citet{DAntona1997} models (Fig.~5) near the 0.3 Myr age isochrone. 
We also plot the 1 Myr isochrone of the models by \citet{Baraffe1998} and
\citet{Chabrier2000} and note that $\sigma$~Ori~IRS1~B is well above this isochrone. 
The differences between both models illustrate the 
problems with theoretical HR diagrams at young ages that the authors of
those models are well aware of \citep{Baraffe2002}.
As an alternative to comparing with evolutionary models, we can compare
$\sigma$~Ori~IRS1~B directly to the components of young the eclipsing binary 2MASSJ05352184-0546085
in the Orion Nebula
whose masses have been directly measured by \citet{Stassun2006}. This star's primary component
of M6.5$\pm$0.5 spectral type has a measured mass of 0.0541 $M_\odot$ while the slightly hotter
secondary component has a mass of 0.0340 $M_\odot$. Since $\sigma$~Ori~IRS1~B is
of later spectral type than either of the components of this eclipsing
binary and is probably of similar age, it seems certain that $\sigma$~Ori~IRS1~B
has a sub-stellar mass below 0.05 $M_\odot$.
Finally, in a strong UV radiation environment, the photospheric temperature 
may be elevated above what an isolated object of otherwise identical properties might
have, which would lead to an overestimate for the mass.

\section{THE EMISSION LINE SPECTRUM OF $\sigma$~ORI~IRS1~B}

The emission line spectrum in the {\it H} and {\it K} bands of $\sigma$~Ori~IRS1~B is shown in
Fig.~2. The spectrum is dominated by strong emission from the Brackett series of HI
and by emission lines of HeI. 
There is, however, no detectable
emission from shock-excited H$_2$ in the S(1) lines, nor is there forbidden
line emission in the 1.644$\mu$m emission line of [FeII], which is often associated with
outflow activity in young stars. 
Emission lines are one of the defining characteristics \citep{Joy1945} of classical T Tauri stars and
in those stars originate from the accretion disk. In the case of $\sigma$~Ori~IRS1~B, however, our
spatially resolved line images show that
the Br$\gamma$ and HeI emission lines do not originate in an accretion disk very close to 
the stellar surface. Rather, the line emission is spatially extended by 95 mas FWHM in the
case of the Br$\gamma$ line, and similar for HeI, while the PSF is 55 mas FWHM. 
The brightest emission is offset towards
the illuminating OB stars as illustrated in Fig.~3. 
On the other hand, the emission lines do not show a pronounced bow-shock morphology,
indicating that we are not seeing IRS1 B
being illuminated side-on. 
This finding agrees with the morphology of the 10 $\mu$m emission found by \citet{vanLoon2003}
who did not see the bow shock structure often found at 10 $\mu$m in side-on Orion proplyds
by \citet{Smith2005}.

While the proplyd IRS1 B was easily detected in H$\alpha$ emission, its surface
brightness falls well short of the prediction by \citet{vanLoon2003} of being
equal to the $\sigma$~Ori~A/B continuum in a 1 nm wavelength interval. 
Calibrating the continuum flux of $\sigma$~Ori~A/B against the
spectrophotometry by \citet{Stone1996} of the HST O9Vp calibration star HD93521, we
get an H$\alpha$ flux for the unresolved H$\alpha$ emission knot from $\sigma$~Ori~IRS1~B of 
4.7~10$^{-15}$~Wm$^{-2}$. The dust that
\citet{vanLoon2003} detected in thermal emission probably absorbs much the H$\alpha$ emission that
would be expected based on extrapolation from the radio spectrum.

\section{DISCUSSION}
The formation of low-mass objects by photo-evaporation of prestellar cores near
massive stars has been discussed by \citet{Whitworth2004}, and our data indicate
that $\sigma$~Ori~IRS1~B appears to show this process in action.
It is interesting to note that the two components of the $\sigma$~Ori~IRS1 system 
show very different spectra. The more luminous component IRS1~A shows a pure photospheric
absorption line spectrum without evidence for dust veiling or emission lines. 
Assuming that the two components are
indeed physically close to each other,
and that IRS1 A is therefore exposed to the same
intense UV radiation field as IRS1 B, it can be concluded that IRS1 A must be without a disk or envelope.
While distant encounter between stars in the dense cores of clusters are efficient at
stripping the outer parts of disks \citep{Pfalzner2005}, the complete lack of any
sign of circumstellar matter around the more massive component A of the IRS1 system
can only be explained by dynamical effects within that system. It can be speculated
that IRS1 A itself might be a very close binary
whose formation may have destroyed the disk, in a scenario similar to those discussed
by \citet{Reipurth2000}.

$\sigma$ Ori~IRS1~B is not the only object undergoing photoevaporation in the
UV radiation field of $\sigma$~Ori~A and B. In a recent paper, 
\citet{Rigliaco2009} showed that the T Tauri star SO587 in the $\sigma$ Ori cluster 
shows emission lines that can best be interpreted in a photoevaporation scenario.

\section{CONCLUSIONS}

We have obtained optical and near-infrared spectroscopy of the binary object
$\sigma$~Ori~IRS1~A and B. 
Component $\sigma$~Ori~IRS1~A shows a pure photospheric absorption line spectrum of M1 type
and is a low mass star. 
Component B of the $\sigma$~Ori~IRS1 binary shows a compound dust and photospheric spectrum.
After accounting for the dust continuum veiling, $\sigma$~Ori~IRS1~B itself can be classified as
a M7 or M8 spectral type and probably is of sub-stellar mass. 
Further, IRS1 B shows a strong emission line
spectrum with the Brackett series of H and HeI lines in the near infrared, and H$\alpha$ in the optical.
The line images in the Br$\gamma$ and HeI (2.06 $\mu$m) lines show that the source of
the line emission is an extended envelope around the star. This envelope of $\sigma$~Ori~IRS1~B 
is clearly being photo-evaporated by the UV radiation field of $\sigma$~Ori~A.

\acknowledgments
We thank Bo Reipurth for helpful discussions.

Some of the data presented herein were obtained at the W.M. Keck Observatory,   
which is operated as a scientific partnership among the California Institute of Technology,
the University of California and NASA.
The Observatory was made possible by the generous financial support of the W.M. Keck Foundation.

\clearpage
\begin{figure}
\figurenum{1}
\includegraphics[scale=1.6,angle=0]{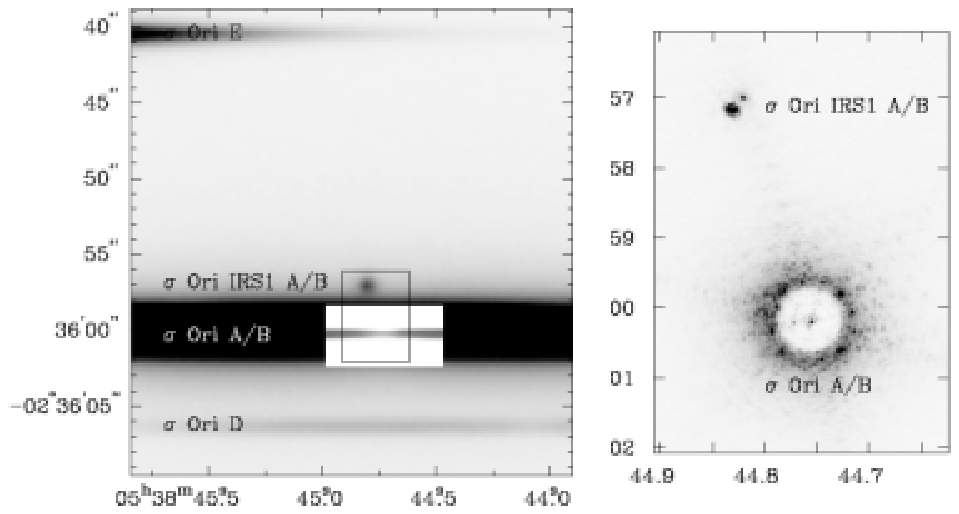}
\caption{
Slit-less spectrum of the $\sigma$ Ori system, pre-filtered and  
centered around H$\alpha$. The brightest continuum is
the binary $\sigma$ Ori A/B with its late O to early B compound spectrum,
with additional contamination from component $\sigma$ Ori C.
The spatial coordinates refer to the wavelength of H$\alpha$.
The H$\alpha$ absorption of the binary $\sigma$ Ori A/B is clearly seen in the higher
signal level window. The proplyd $\sigma$ Ori IRS1 B is the object detected
only by its H$\alpha$ emission line to the north of the bright binary.
The image to the right is the
Keck NIRC-2 adaptive optics image of the $\sigma$ Ori system in
the Br$\gamma$+continuum filter, covering the
area indicated by the box in the H$\alpha$ image.
The main components
$\sigma$ Ori A/B are visible behind the partly transparent occulting mask.
}
\end{figure}

\clearpage
\begin{figure}
\figurenum{2}
\includegraphics[scale=0.9,angle=0]{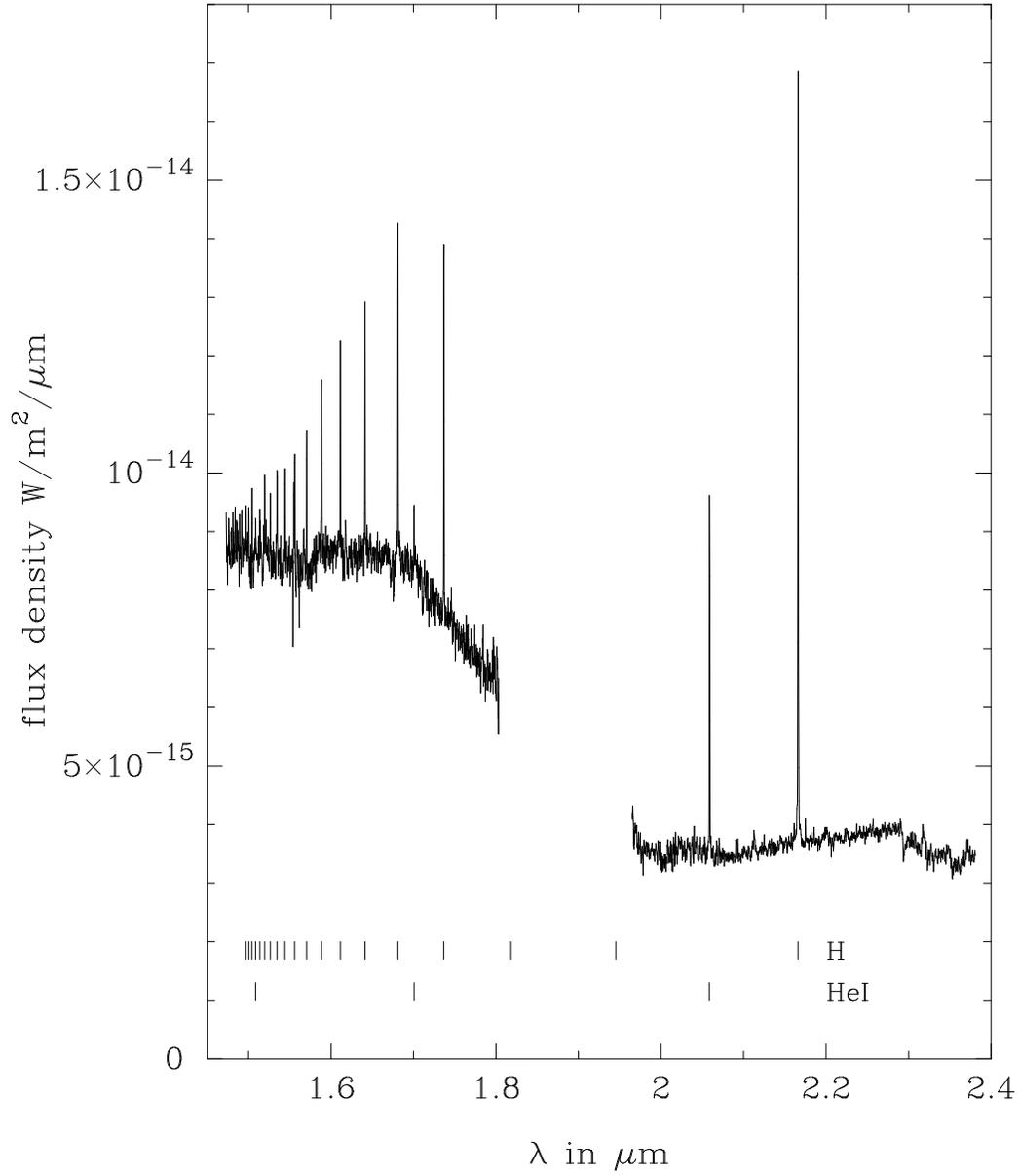}
\caption{
Spectrum of $\sigma$Ori IRS1 B in the {\it H} and {\it K} bands, showing emission
lines of H and HeI superposed over a continuum spectrum composed
of a M7.5 photosphere and dust continuum veiling.
}
\end{figure}

\clearpage
\begin{figure}
\figurenum{3}
\includegraphics[scale=0.9,angle=0]{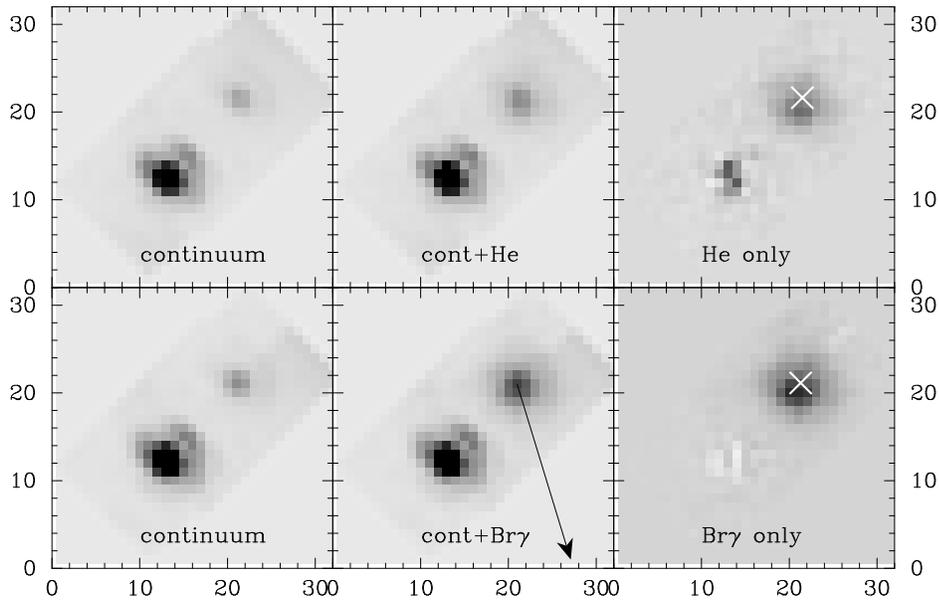}
\caption{
Extracted adjacent continuum, line plus underlying continuum, and continuum-subtracted line images
of $\sigma$ Ori IRS1 A and B in the emission lines of HeI (2.059$\mu$m) and H Br$\gamma$ (2.166$\mu$m).
Tickmarks are in units of the OSIRIS 20 mas pixels.
The arrow indicates the direction towards the illuminating UV source $\sigma$Ori A/B.
The X symbol in the continuum-subtracted images indicates the position of the star.
It should be noted that $\sigma$Ori IRS1 A is unresolved and that all the structure
seen around it is due to the PSF. 
}
\end{figure}

\clearpage
\begin{figure}
\figurenum{4}
\includegraphics[scale=0.8,angle=0]{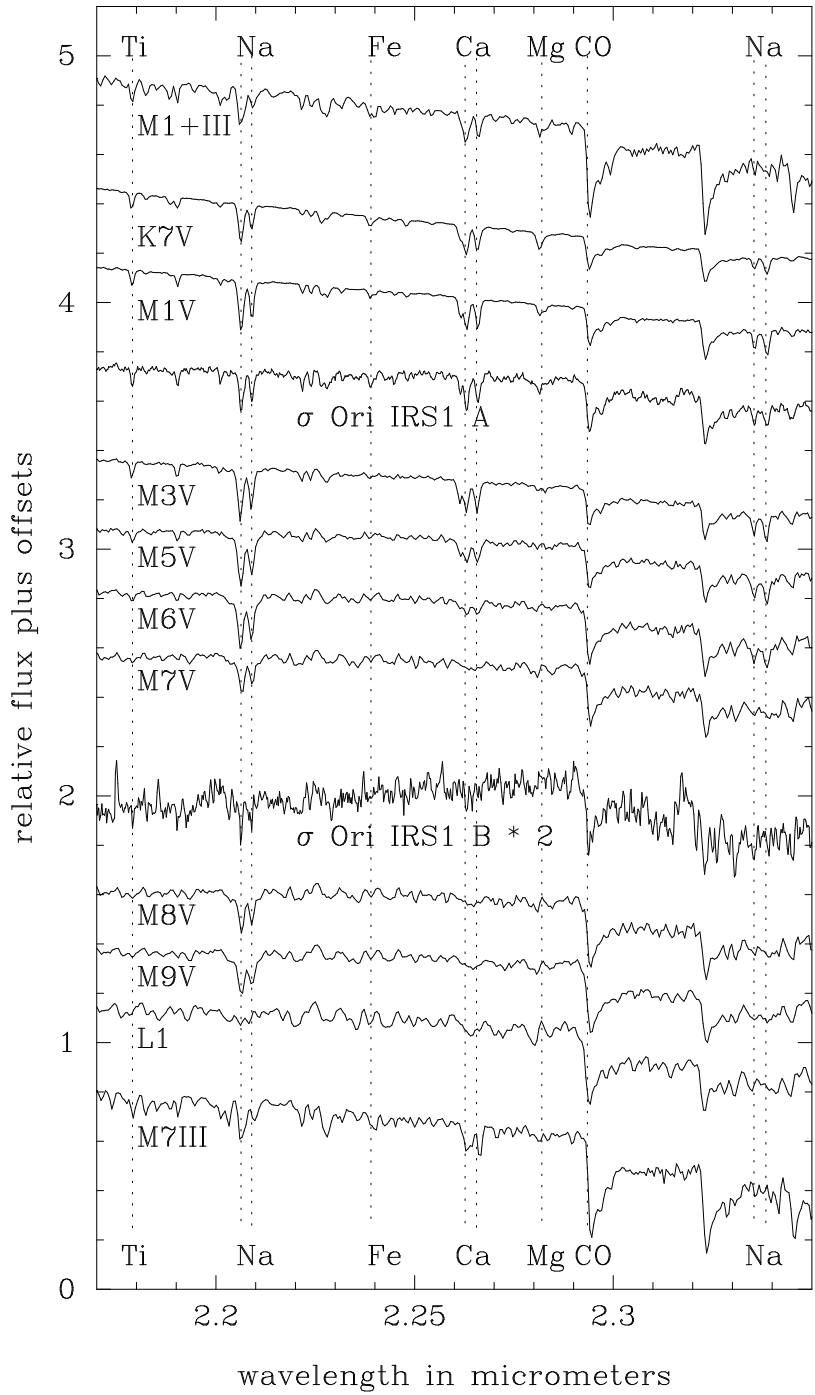}
\caption{
Continuum spectra of $\sigma$Ori IRS1 A and IRS1 B in the {\it K} window
and comparison spectra from the IRTF spectral library \citep{Cushing2005}.
}
\end{figure}

\clearpage
\begin{figure}
\figurenum{5}
\includegraphics[scale=0.8,angle=0]{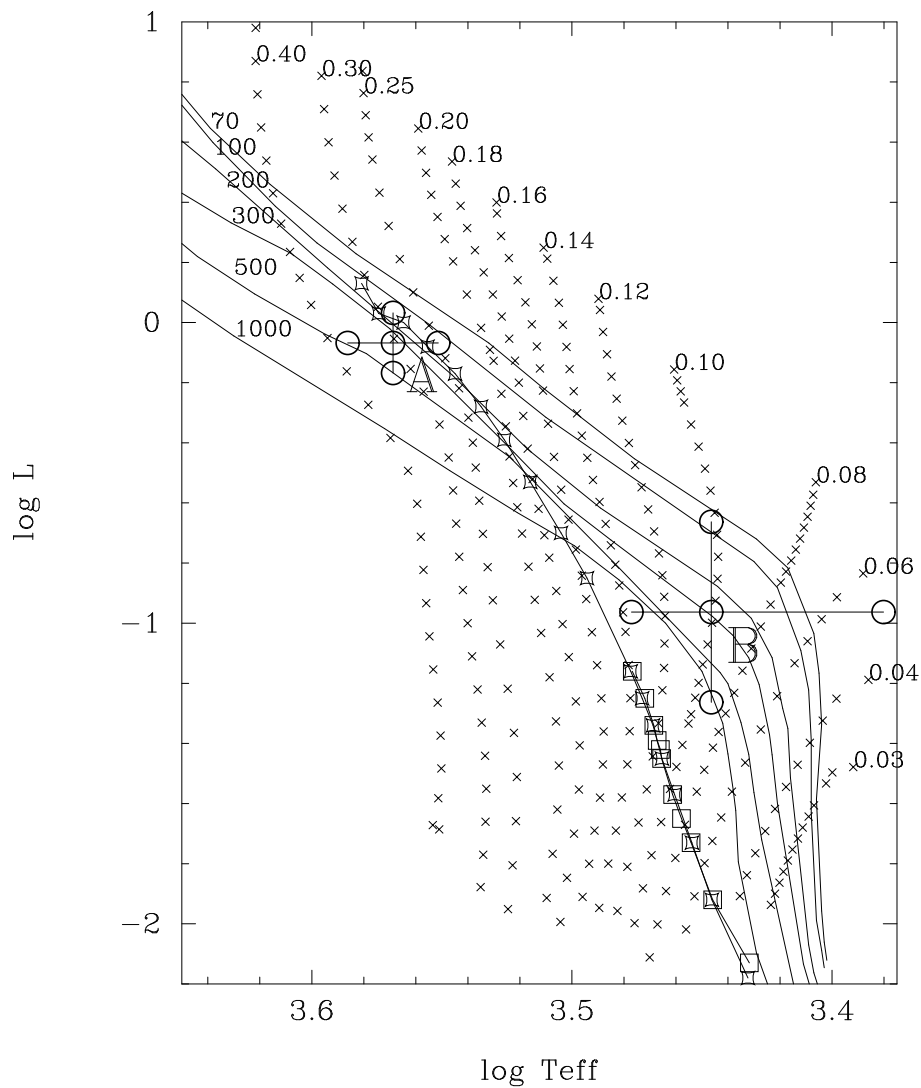}
\caption{
Pre-main-sequence evolutionary tracks from \citet{DAntona1997}
(x symbols).
Their Isochrones are labeled in units of 1000 yrs.
The 1 Myr isochrone from \citet{Baraffe1998} and \citet{Chabrier2000}
is indicated by open squares.
The HR diagram loci of
$\sigma$ Ori IRS1 A and B are indicated. 
The error bars indicate the errors in spectral type determination
and photometric errors.
}
\end{figure}


\clearpage

\begin{thebibliography}{}

\bibitem[Baraffe et al.(1998)]{Baraffe1998}
Baraffe, I., Chabrier, G., Allard, F., \& Hauschildt, P. H. 1998, \aap, 337, 403

\bibitem[Baraffe et al.(2002)]{Baraffe2002}
Baraffe, I., Chabrier, G., Allard, F., \& Hauschildt, P. H. 2002, \aap, 382, 563

\bibitem[Bolton (1974)]{Bolton1974}
Bolton, C. T. 1974, \apj, 192, L7

\bibitem[Bouy et al.(2009)]{Bouy2009}
Bouy, H., Hu{\'e}lamo, N., Mart{\'{\i}}n, E.~L., Marchis, F., Barrado y Navascu{\'e}s, D., Kolb, J., 
Marchetti, E., Petr-Gotzens, M. G., Sterzik, M., Ivanov, V. D., K{\"o}hler, R., \& N{\"u}rnberger 2009, \aap, 493, 931

\bibitem[Chabrier et al.(2000)]{Chabrier2000}
Chabrier, G., Baraffe, I., Allard, F., \& Hauschildt, P. 2000, \apj, 542, 464


\bibitem[Caballero (2007b)]{Caballero2007b}
Caballero, J. A. 2007, AN 328, 1

\bibitem[Caballero (2007a)]{Caballero2007a}
Caballero, J. A. 2007, \aap, 466, 917

\bibitem[Cohen, Wheaton, \& Megeath (2003)]{Cohen2003}
Cohen, M., Wheaton, Wm. A., \& Megeath, S. T. 2003, \aj, 126, 1090

\bibitem[Cushing, Rayner, \& Vacca (2005)]{Cushing2005}
Cushing, M. C., Rayner, J. T., \& Vacca, W. D. 2005, \apj, 623, 1115

\bibitem[D'Antona \& Mazzitelli (1994)]{DAntona1994}
D'Antona, F. \& Mazzitelli, I. 1994, \apjs, 90, 467

\bibitem[D'Antona \& Mazzitelli (1997)]{DAntona1997}
D'Antona, F. \& Mazzitelli, I. 1997, Mem. Soc. Astr. Ital., 68, 807

\bibitem[Drake (1990)]{Drake1990}
Drake, S. A. 1990, \aj, 100, 572

\bibitem[Garrison (1967)]{Garrison1967}
Garrison, R. F., 1967, \pasp, 79, 433

\bibitem[Greene \& Lada (1996)]{Greene1996}
Greene, T. P. \& Lada, C. J. 1996, \aj, 112,2184

\bibitem[Heintz (1997)]{Heintz1997}
Heintz, W. D. 1997, \apjs, 111, 335

\bibitem[Hodapp \& Deane (1993)]{Hodapp1993}
Hodapp, K.-W., \& Deane, J. 1993, \apjs, 88, 119

\bibitem[Johnson (1965)]{Johnson1965}
Johnson, H. L. 1965, \apj, 141, 170

\bibitem[Joy (1945)]{Joy1945}
Joy, A. H. 1945, \apj, 102, 168

\bibitem[Kleinmann \& Hall (1986)]{Kleinmann1986}
Kleinmann, S. G., \& Hall, D. N. B. 1986, \apjs, 62, 501

\bibitem[Larkin et al.(2006)]{Larkin2006}
Larkin, J. et al. 2006, New Astr. Reviews, 50, 362

\bibitem[Luhman et al.(2003)]{Luhman2003}
Luhman, K. L., Stauffer, J. R., Muench, A. A., Rieke, G. H.,
Lada, E. A., Bouvier, J., \& Lada, C. J. 2003, \apj, 593, 1093

\bibitem[Malkov et al.(1997)]{Malkov1997}
Malkov, O. Yu., Piskunov, A. E., \& Shpil'kina, D. A. 1997, \aap, 320, 79

\bibitem[Massey et al.(2005)]{Massey2005}
Massey, P., Puls, J., Pauldrach, A. W. A., Bresolin, F., Kudritzki, R. P., \& Simon, T. 2005,
\apj, 627, 477

\bibitem[Pfalzner, Umbreit, \& Henning (2005)]{Pfalzner2005}
Pfalzner, S., Umbreit, S., \& Henning, Th. 2005, \apj, 629, 526

\bibitem[Reipurth et al.(1998)]{Reipurth1998}
Reipurth, B., Bally, J., Fesen, R. A., \& Devine, D. 1998, Nature, 396, 343

\bibitem[Reipurth (2000)]{Reipurth2000}
Reipurth, B. 2000, \aj, 120, 3177

\bibitem[Rigliaco et al.(2009)]{Rigliaco2009}
Rigliaco, E., Natta, A., Randich, S., \& Sacco, G. 2009, \aap, in press (arXiv:0902.0457v1)

\bibitem[Sherry et al.(2008)]{Sherry2008}
Sherry, W. H., Walter, F. M. , Wolk, S. J. \& Adams, N. R. 2008, \aj, 135, 1616

\bibitem[Skinner et al.(2008)]{Skinner2008}
Skinner, S. L., Sokal, K. R., Cohen, D. H., Gagn{\'e}, M., Owocki, S. P., \& Townsend, R. D. 2008,
\apj, 683, 796

\bibitem[Skrutskie et al.(2006)]{skr06}
Skrutskie, M. F., Cutri, R. M., Stiening, R., Weinberg, M. D., Schneider, S.,
Carpenter, J. M., Beichman, C., Capps, R., Chester, T., Elias, J., Huchra, J.,
Liebert, J., Lonsdale, C., Monet, D. G., Price, S., Seitzer, P., Jarrett, T.,
Kirkpatrick, J. D., Gizis, J., Howard, E., Evans, T., Fowler, J., Fullmer, L., 
Hurt, R., Light, R., Kopan, E. L., Marsh, K. A., McCallon, H. L., Tam, R.,
Van Dyk, S., \& Wheelock, S. 2006, \aj, 131, 1163

\bibitem[Smith (2005)]{Smith2005}
Smith, N., Bally, J., Shuping, R. Y., Morris, M., \& Kassis, M. 2005, \aj, 130, 1763

\bibitem[Stassun, Mathieu, \& Valenti (2006)]{Stassun2006}
Stassun, K. G., Mathieu, R. D., \& Valenti, J. A. 2006, Nature, 440, 311

\bibitem[Stone (1996)]{Stone1996}
Stone, R. P. S. 1996, \apjs, 107, 423

\bibitem[Struve (1837)]{Struve1837}
Struve, F. G. W. 1837, Astronomische Nachrichten, 14, 249

\bibitem[Uehara et al.(2004)]{Uehara2004} Uehara, M., et al.\ 
2004, \procspie, 5492, 661 

\bibitem[van Loon \& Oliveira (2003)]{vanLoon2003}
van Loon, J. Th., Oliveira, J. M. 2003, \aap, 405, L33

\bibitem[Walter et al.(2008)]{Walter2008}
Walter, F. M., Sherry, W. H., Wolk, S. J., \& Adams, N. R. 2008,
in Handbook of Star Forming Regions, Vol. I, Astronomical Society of the Pacific, 
Bo Reipurth, ed., p. 732.

\bibitem[Whitworth \& Zinnecker (2004)]{Whitworth2004}
Whitworth, A. P. \& Zinnecker, H. 2004, \aap, 427, 299

\bibitem[Zapatero Osorio et al.(2008)]{Zapatero2008}
Zapatero Osorio, M. R., B{\'e}jar, V. J. S., Bihain, G., Mart{\'i}n, E. L., Rebolo, R., Vill{\'o}-P{\'e}rez, I.,
D{\'i}az-S{\'a}nchez, A., P{\'e}rez Garrido, A., Caballero, J. A., Henning, T., Mundt, R., Barrado y Navascu{\'e}s, D.,
\& Bailer-Jones, C. A. L. 2008, \aap, 477, 895

\end{thebibliography}
\end{document}